\documentclass[
    ,final            
  ,numberedheadings 
  ]
  {aipproc}

\layoutstyle{6x9}

\begin{document}

\title{Electromagnetic Productions of $K\Lambda$ and $K\Sigma$ on the 
    Nucleons}


\author{T. Mart}{
  address={Departemen Fisika, FMIPA, Universitas Indonesia, 
Depok, 16424, Indonesia}
}

\begin{abstract}
 We briefly review the progress and problems in the electromagnetic 
 production of $K\Lambda$ on the nucleon. The problem of the
 data discrepancy in this channel as well as the corresponding physics consequence
 are highlighted. We also discuss the effect of the new 
 beam-recoil polarization data $C_x$
 and $C_z$ on our analysis.
 For this purpose we use the isobar model Kaon-Maid
 and a recent multipoles model that can describe recent experimental
 data.  We also present a new multipoles model for the 
 $K\Sigma$ channels to complete our analysis.
\end{abstract}

\maketitle

\section{Introduction}
One of the most important goals of nuclear and particle physics
is a unified understanding of the baryon-baryon interaction.
However, 
unlike in the case of the nucleon-nucleon interactions,  
our knowledge on the hyperon-nucleon interactions is far from complete.
The lack of hyperon beam or target becomes the main reason of this
difficulty. Thus, one needs indirect reactions to study this strange
particles. On the other hand, the strange quark in this 
particle generates another degree of
freedom and, therefore, gives additional information
not available from the nucleon-nucleon scattering processes.
As a consequence, investigations of the 
strange particles remain an interesting research topic nowadays.
This is also supported by the fact that hypernuclear studies relies 
heavily on the available information on the hyperon-nucleon interactions.
To this end, the associated production of strange particles is
very helpful, both as a source of information on the hyperon-nucleon
interaction and as the elementary operator that describes the process
at the elementary level.
The electromagnetic production of kaon on the nucleon is one of 
the commonly used reactions for this purpose. Both virtual and real photons
can be used. However, since the real photon is theoretically much
simpler than the virtual one, we will limit the following discussion
to the photoproduction process. 

\section{Electromagnetic Productions of $K\Lambda$}
In what follows, we shall consider two phenomenological models
based on the Feynman and multipoles techniques, i.e., the Kaon-Maid
model~\cite{kaon-maid,Mart:1999ed} 
and the recent multipole approach given in Ref.~\cite{Mart:2006dk}. 
In the former, tree-level Feynman diagrams have been used 
to reproduce all available $K^+ \Lambda$, $K^+ \Sigma^0$ and $K^0 \Sigma^+$
photoproduction observables. The background terms contain the standard
$s$-, $u$-, and $t$-channel along with a contact
term, which is required to restore gauge invariance after hadronic
form factors had been introduced \cite{hbmf}. Furthermore, four 
nucleon resonances, the $S_{11}$(1650), $P_{11}$(1710), $P_{13}(1720)$,
and the ``missing resonance'' $D_{13}(1895)$ have been also included 
in this model. 
For $K \Sigma$ production further contributions from the
$S_{31}$(1900) and $P_{31}$(1910) $\Delta$ resonances were added.
Note that, Kaon-Maid was fitted to old and previous version of SAPHIR data~\cite{SP98}.
An interactive version of this model is available through 
internet~\cite{kaon-maid-homepage}. 

The multipole model utilizes
the same background terms, whereas the resonance parts are assumed 
to have the Breit-Wigner form~\cite{Tiator:2003uu}
\begin{eqnarray}
  \label{eq:em_multipole}
  A_{\ell\pm}^R(W) = {\bar A}_{\ell\pm}^R  c_{KY} \frac{f_{\gamma R}(W)
    \Gamma_{\rm tot}(W) M_R\, f_{K R}(W)}{M_R^2-W^2-iM_R\Gamma_{\rm tot}(W)} e^{i\phi},~ 
  \label{eq:m_multipole}
\end{eqnarray}
where $W$ the total c.m. energy, $c_{KY}$ the isospin factor, 
$f_{KR}$ the conventional Breit-Wigner factor describing 
the decay of a resonance $R$ with a total width
$\Gamma_{\rm tot}(W)$ and physical mass $M_R$, $f_{\gamma R}$ 
the $\gamma NR$ vertex factor, and $\phi$ the phase angle. The
model was fitted to the combinations of the recent 
SAPHIR~\cite{Glander:2003jw}, CLAS~\cite{Bradford:2005pt},
and LEPS~\cite{Sumihama:2005er,Kohri:2006yx} data. In spite of
their unprecedented high qualities, these new data sets, 
however, reveal a lack of consistency at the forward and 
backward kaon angles. This problem hinders the reliable extraction 
of the resonance parameters, which could lead to different 
conclusions on the extracted ``missing resonances''.

\subsection{Differences between CLAS and SAPHIR Data}

\subsubsection{Statistical Differences}
\begin{figure}
  \includegraphics[height=.28\textheight]{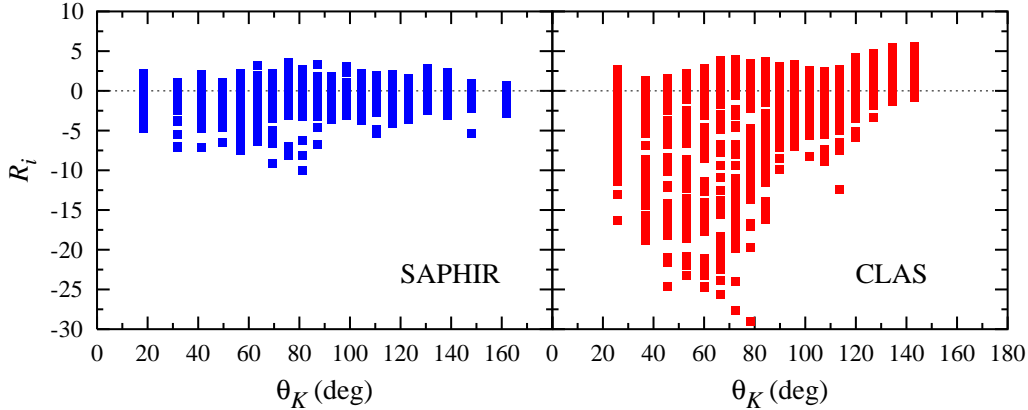}
  \caption{Deviations of the predictions of the Kaon-Maid model from the  
    SAPHIR and CLAS experimental data points as a function of the kaon 
    c.m. angle. Note that $R_i$ is defined in Eq.~(\protect\ref{deviation}).
    \label{fig:ri}}
\end{figure}

Reference \cite{Bydzovsky:2006wy} has studied the statistical
properties of both CLAS and SAPHIR data in a great detail 
by using four different
isobar models. In general it is found that, compared to the other
three models, the Kaon-Maid model provides a better description 
of the presently existing data. Nevertheless, the agreement with
the SAPHIR data is more remarkable than with other data,
 which is indicated by the
fact that the SAPHIR data are scattered closer to $R_i=0$
compared to the CLAS ones (see Fig.~\ref{fig:ri}), 
where $R_i$ is
the relative deviation of each data point, defined by
\begin{eqnarray}
R_i=\frac{\sigma_i^{\rm exp}-\sigma^{\rm th}(E_i,\theta_i)}
{\Delta\sigma_i^{\rm stat}} ~.
\label{deviation}
\end{eqnarray}
Interestingly, if we analyze this agreement more closely by using
the statistical parameter $z_1$, then a different phenomenon appears.
The parameter is defined as
\begin{eqnarray}
z_1 = \sqrt{N-1}\frac{\langle R\rangle}{\sqrt{\langle(\Delta R)^2\rangle}} ,
\label{z_1stat}
\end{eqnarray}
where $N$ is the number of data points and 
$\langle(\Delta R)^2\rangle=\langle R^2\rangle -\langle R\rangle ^2$ 
indicates the square of the variance of the normal distribution of $R_i$. 
Provided that the data are randomly scattered around the theoretical 
values with this variance, the hypothesis that the true value of 
the mean $\langle R\rangle$ equals zero (the null hypothesis) can 
be rejected with a confidence level of $\alpha$
if $|z_1|>z_{\alpha/2}$, where the critical value $z_{\alpha/2}$ = 1.96
and 2.58 for the confidence level of 5\% and 1\%, respectively \cite{StatMan}.

As shown in Ref.~\cite{Bydzovsky:2006wy},
the use of SAPHIR data in Kaon-Maid model yields $|z_1|=11.7$, whereas the 
use of CLAS data in the same model results in  $|z_1|=1.41$. 
Focusing only on the forward-direction data
does not change this result. This leads
to the conclusion that if we reject the null hypothesis, then 
there is a large probability that we are wrong. In other words, the 
Kaon-Maid model is more consistent with the CLAS data.

Recent analyses have also indicated that there could be a global scaling
factor between the CLAS and SAPHIR data. To determine this factor, 
Ref.~\cite{Bydzovsky:2006wy} defined the quantity $s$ through
\begin{eqnarray}
  \label{eq:scaling}
  \chi_0^2 &=& \sum_{i=1}^{N}\left(
    \frac{s\sigma_i^{\rm exp}-\sigma^{\rm th}
      (E_i,\theta_i)}{\Delta\sigma_i^{\rm stat}}
  \right)^2~,
\end{eqnarray}
and minimized the $\chi_0^2$ by using the SAPHIR data, 
where $\sigma^{\rm th}$ is obtained from 
the specific isobar model that had been previously fitted to the
CLAS data. For the full data set it is found that $s=1.13$ and
for the forward data set the best fit yields $s=1.15$~\cite{Bydzovsky:2006wy}.
These findings indicate that an increase of the SAPHIR data by a factor
of $13\%$ -- $15\%$ would improve the agreement between the two
data sets. These values are, however, smaller than the previously suggested
scaling factor of $\sim 4/3$~\cite{Bradford:2005pt}. 

\subsubsection{The Physics Consequences}
\begin{figure}
\begin{minipage}[htb]{75mm}
\includegraphics[height=.23\textheight]{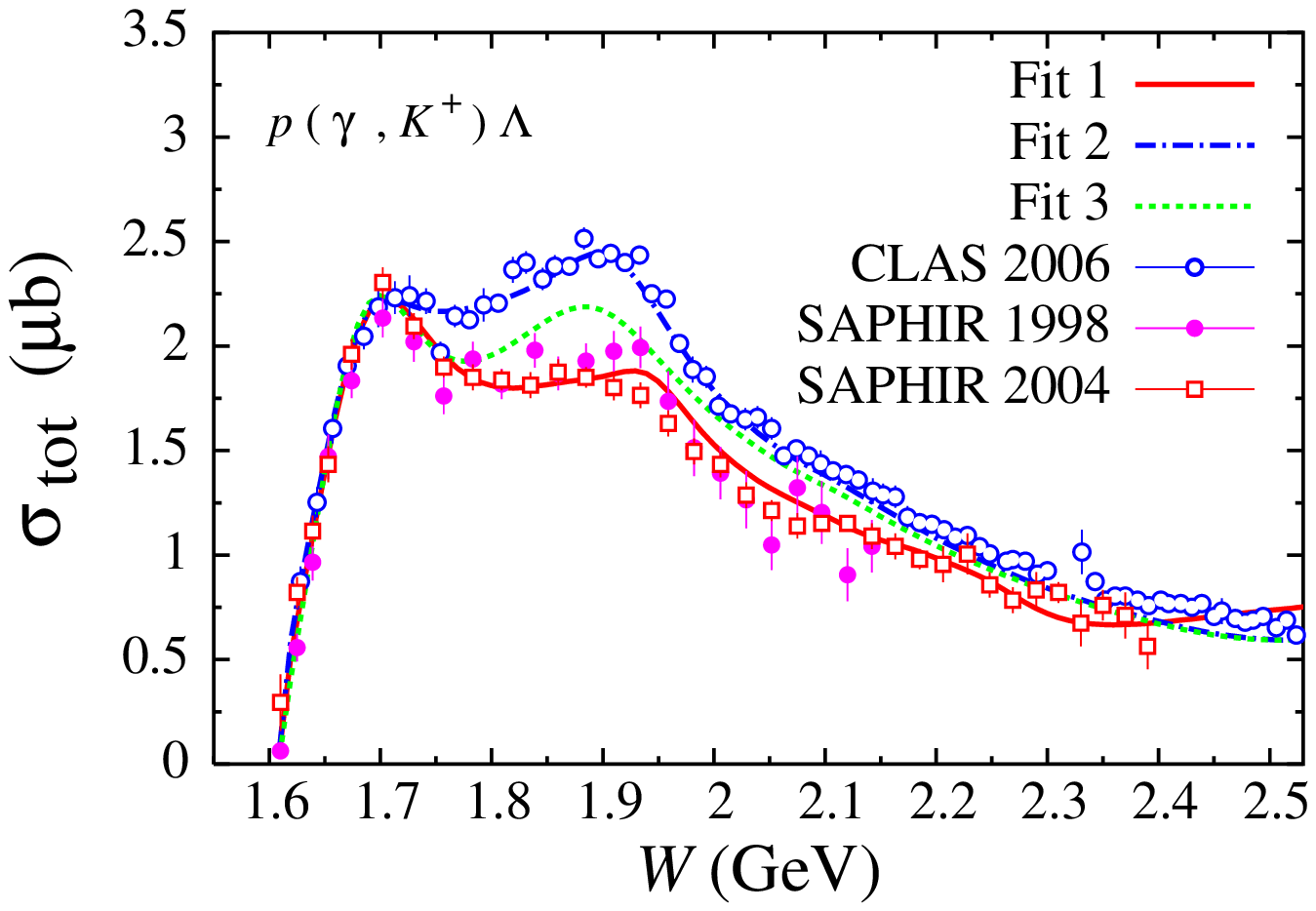}
\end{minipage}
\hspace{\fill}
\begin{minipage}[htb]{75mm}
  \includegraphics[height=.21\textheight]{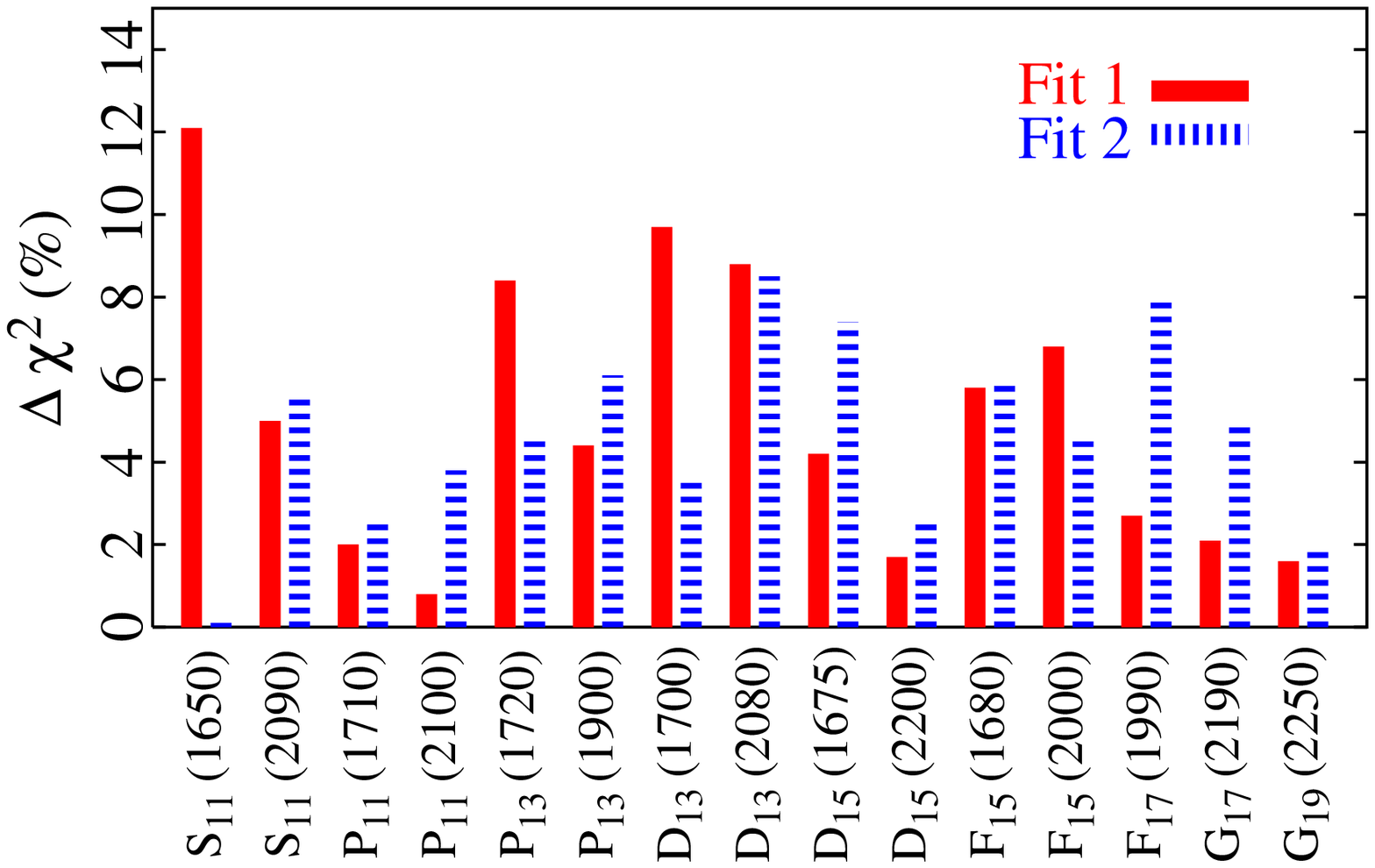}
\end{minipage}
  \caption{{\bf (Left)} Comparison between the calculated total cross sections
    with experimental data, which clearly shows the discrepancy problem
    between the CLAS and SAPHIR data~\cite{Mart:2006dk}. 
    {\bf (Right)} The importance of individual resonances in the multipole 
    models that fit to SAPHIR (Fit 1) and CLAS (Fit 2) 
    data~\cite{Mart:2006dk}. Note that 
    $\Delta \chi^2 = {\left|\chi^2_{\rm All}-\chi^2_{{\rm All}-N^*}\right|}/{\chi^2_{\rm All}}
  \times 100\,\%$, whereas Fit 3 is obtained by using all data sets.
    \label{fig:strength}}
\end{figure}

The problem of the lack of mutual consistency between the SAPHIR and CLAS data
has certainly some physics consequences. 
The use of SAPHIR and CLAS data, 
individually or simultaneously, leads to quite different resonance parameters 
which, therefore, could lead to different conclusions on the ``missing resonances''. 
This is shown in Fig.~\ref{fig:strength}. 
Fitting to the SAPHIR data (denoted by Fit 1 in the figure) indicates that 
the $S_{11}(1650)$, $P_{13}(1720)$, $D_{13}(1700)$, 
$D_{13}(2080)$, $F_{15}(1680)$, and 
$F_{15}(2000)$ resonances are required, while fitting to the 
of CLAS (Fit 2) data leads alternatively to the $P_{13}(1900)$, $D_{13}(2080)$, 
$D_{15}(1675)$, $F_{15}(1680)$, and $F_{17}(1990)$ resonances.
Fitting both data sets simultaneously (Fit 3) yields a compromise result and 
changes this conclusion which indicates that the corresponding result 
is neither consistent with Fit 1 nor with Fit 2. 

\begin{figure}[b]
  \includegraphics[height=.26\textheight]{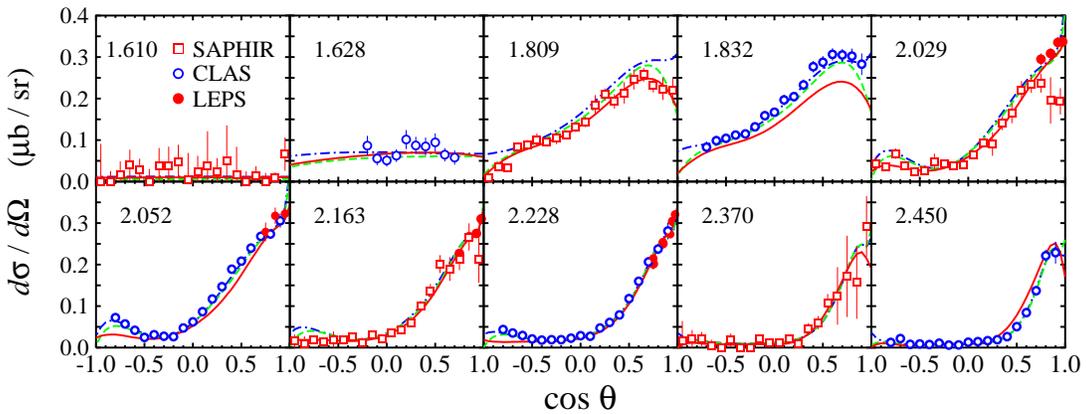}
  \caption{Comparison between the calculated differential cross sections
    obtained from a multipole model~\cite{Mart:2006dk}
    with some selected experimental data. Notation for the curves is 
    as in Fig.~\ref{fig:strength}. 
    \label{fig:dcs_kl}}
\end{figure}

Although yielding different results in most cases 
(see Fig.~\ref{fig:dcs_kl}) 
both SAPHIR 
and CLAS data indicate that the second peak in the total cross 
sections at $W\sim 1900$ MeV, 
shown in the left panel of Fig.~\ref{fig:strength}, 
originates from the $D_{13}(2080)$ resonance.
By refitting the Kaon-Maid model to the CLAS and SAPHIR
data individually, it is shown that the extracted masses of 
the missing resonance $D_{13}(1895)$ differ only by 11 
MeV~\cite{Bydzovsky:2006wy}. The same situation is also
found in the multipole model~\cite{Mart:2006dk}.
This is clearly demonstrated in Table~\ref{tab:missing}.
The extracted widths, however, vary from 165 to 570 MeV. 

The Gerasimov-Drell-Hearn (GDH) sum rule also provides another tool
to investigate the physics difference between the CLAS and SAPHIR data.
The sum rule relates the anomalous magnetic 
moment of the nucleon $\kappa_N$ to the difference of its
polarized total photoabsorption cross sections \cite{gdh}. 
Since there has been no available measurement, these  cross sections 
must be predicted 
from a reliable model which fits all
existing unpolarized experimental data. The two models (Fit 1 and Fit 2)
described above can be used for this purpose. It is found that the two data
sets yield quite different contributions~\cite{Mart:2008ik}. 
The predicted contribution of Fit 1 is much closer to that of the Kaon-Maid, 
indicating the consistency of the new
SAPHIR data \cite{Glander:2003jw} to the old ones \cite{SP98}.
The model that fits the CLAS differential cross section data
(Fit 2) tends to eliminate the contribution of kaon-hyperon
final states to the GDH sum rule.  

\begin{table}[t]
  \begin{tabular}[c]{lcccccc}
    \hline
    & \multicolumn{3}{c}{Kaon-Maid} && \multicolumn{2}{c}{Multipole}\\
    \cline{2-4}\cline{6-7}\\[-2ex]
    & Original \cite{SP98} & SAPHIR~\cite{Glander:2003jw} & CLAS~\cite{Bradford:2005pt}
    && SAPHIR~\cite{Glander:2003jw} & CLAS~\cite{Bradford:2005pt}\\
    \hline
    $M$ (GeV)      & $1.895\pm 0.004$ & $1.938\pm 0.004$ & $1.927\pm 0.003$ && $1.936\pm 0.010$ & $1.915\pm 0.004$\\
    $\Gamma$ (GeV) & $0.372\pm 0.029$ & $0.233\pm 0.008$ & $0.570\pm 0.019$ && $0.301\pm 0.022$ & $0.165\pm 0.008$\\
    \hline
  \end{tabular}
\caption{The values of mass ($M$) and width ($\Gamma$) of
    the missing $D_{13}$ resonance extracted from Kaon-Maid 
    using the three different experimental data~\protect\cite{Bydzovsky:2006wy}
    and from a multipole model using SAPHIR and CLAS data \cite{Mart:2006dk}.}
\label{tab:missing}
\end{table}

\subsection{Influence of the New $C_x$ and $C_z$ Data}
\begin{figure}
  \includegraphics[height=.28\textheight]{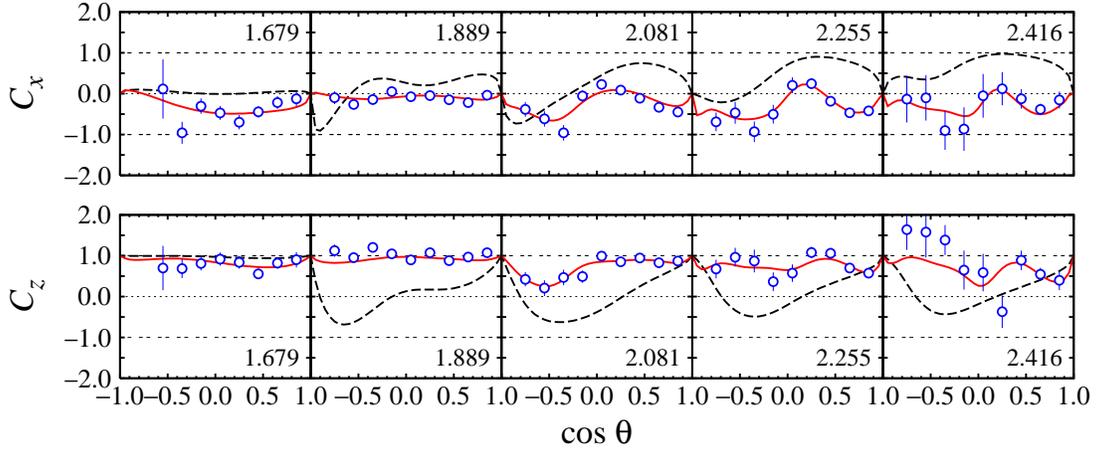}
  \caption{Sample of the beam-recoil polarization observables $C_x$ and $C_z$ 
      for the reaction $\vec{\gamma} p\to K^+\vec{\Lambda}$ 
      plotted as a function of the kaon scattering angle. Experimental data
      are taken from Ref.~\cite{Bradford:2006ba}. The corresponding total c.m. energy $W$
      is shown in each panel. Dashed curves show the
      prediction of Kaon-Maid, solid curves demonstrate the result
      of the multipole model after including the $C_x$ and $C_z$ data. 
      \label{fig:cx}}
\end{figure}

\begin{figure}[b]
  \includegraphics[height=.25\textheight]{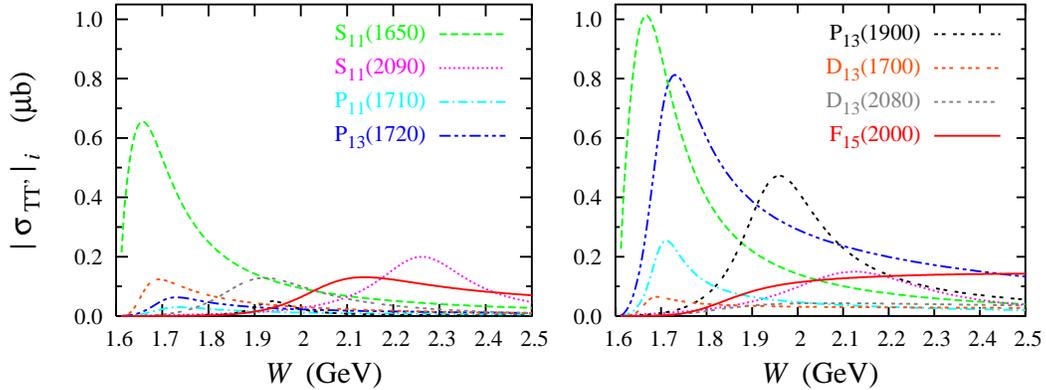}
  \caption{The individual contribution of several 
      important resonances to the absolute value of the 
      total cross section $\sigma_{\rm TT'}$ 
      before (left panel) and after (right panel) the inclusion of 
      the beam-recoil polarization observables $C_x$ and $C_z$. 
      Note that contributions from other resonances are small and, 
      therefore, are not shown in this figure for the sake of clarity.
      \label{fig:sig_tt}}
\end{figure}

Recently, a set of the beam-recoil polarization observables data, $C_x$ and 
$C_z$, has been released by  the CLAS collaboration~\cite{Bradford:2006ba}.
These data indicate that the $\Lambda$ polarization is 
predominantly in the direction of the spin of the 
incoming photon, independent of the c.m. energy or 
the kaon scattering angle (see Fig.~\ref{fig:cx}). 
Recent analyses found that 
these data seems to be difficult to explain. Clearly, 
it is interesting to include these data in our analysis, 
as well as to investigate the effects of the data 
inclusion~\cite{Mart:2008ik}. After including these
data it is found that the total cross sections $\sigma_{\rm TT'}$ show
less structures. This indicates that the CLAS $C_x$ and $C_z$ data select
certain resonances as the important ones. To investigate this phenomenon, in
Fig.~\ref{fig:sig_tt} we plot contributions of 
several important resonances to the total cross section 
$\sigma_{\rm TT'}$ before and after the inclusion of the 
$C_x$ and $C_z$ data~\cite{Mart:2008ik}. It is obvious from this figure that 
the inclusion emphasizes the roles of
the $S_{11}(1650)$, $P_{11}(1710)$, $P_{13}(1720)$, and $P_{13}(1900)$
resonances, which corroborates
the finding of the authors of Ref.~\cite{Anisovich:2007bq}

\section{Electromagnetic Productions of $K\Sigma$}
Photoproductions of $K\Sigma$ are of interest because existing 
models that can nicely reproduce the $K^+\Sigma^0$ data could 
overestimate the charged $\Sigma$ data by almost two orders of 
magnitude~\cite{Mart:1995wu}. In these channels the amplitudes
$F_i$, can be expressed in terms of three independent isospin
amplitudes, i.e. $A^{(0)}$ for the isoscalar photon, $A^{(1/2)}$ and $A^{(3/2)}$
for the isovector photon with total isospin of the $KY$ system $I=1/2$ and
$I=3/2$, respectively. For comparison with the results of previous calculations,
as well as with the PDG values~\cite{yao:2006}, it is also useful to define the proton
$_pA^{(1/2)}$ and neutron $_nA^{(1/2)}$ helicity photon couplings with total isospin 1/2,
\begin{eqnarray}
  \label{eq:pn_amplitudes}
  _pA^{(1/2)} ~=~ A^{(0)}+{\textstyle \frac{1}{3}}\, A^{(1/2)} ~~,~~~~
  _nA^{(1/2)} ~=~ A^{(0)}-{\textstyle \frac{1}{3}}\, A^{(1/2)} ~.
\end{eqnarray}
Using this notation, the CGLN amplitudes for the four
physical channels of kaon photoproduction can be written as
\begin{eqnarray}
  \label{eq:amplitude_four_channels}
  A(\gamma p\to K^+\Sigma^0) &=& _pA^{(1/2)}+{\textstyle \frac{2}{3}}\, A^{(3/2)}~,\\
  A(\gamma n\to K^0\Sigma^0) &=& - _nA^{(1/2)}+{\textstyle \frac{2}{3}}\, A^{(3/2)}~,\\
  A(\gamma p\to K^0\Sigma^+) &=& \sqrt{2}\left[ _pA^{(1/2)}-{\textstyle \frac{1}{3}}\, A^{(3/2)}\right]~,\\
  A(\gamma n\to K^+\Sigma^-) &=& \sqrt{2}\left[ _nA^{(1/2)}+{\textstyle \frac{1}{3}}\, A^{(3/2)}\right]~.
\end{eqnarray}
All observables are calculated from these amplitudes. 

\begin{figure}[!t]
  \includegraphics[height=.5\textheight]{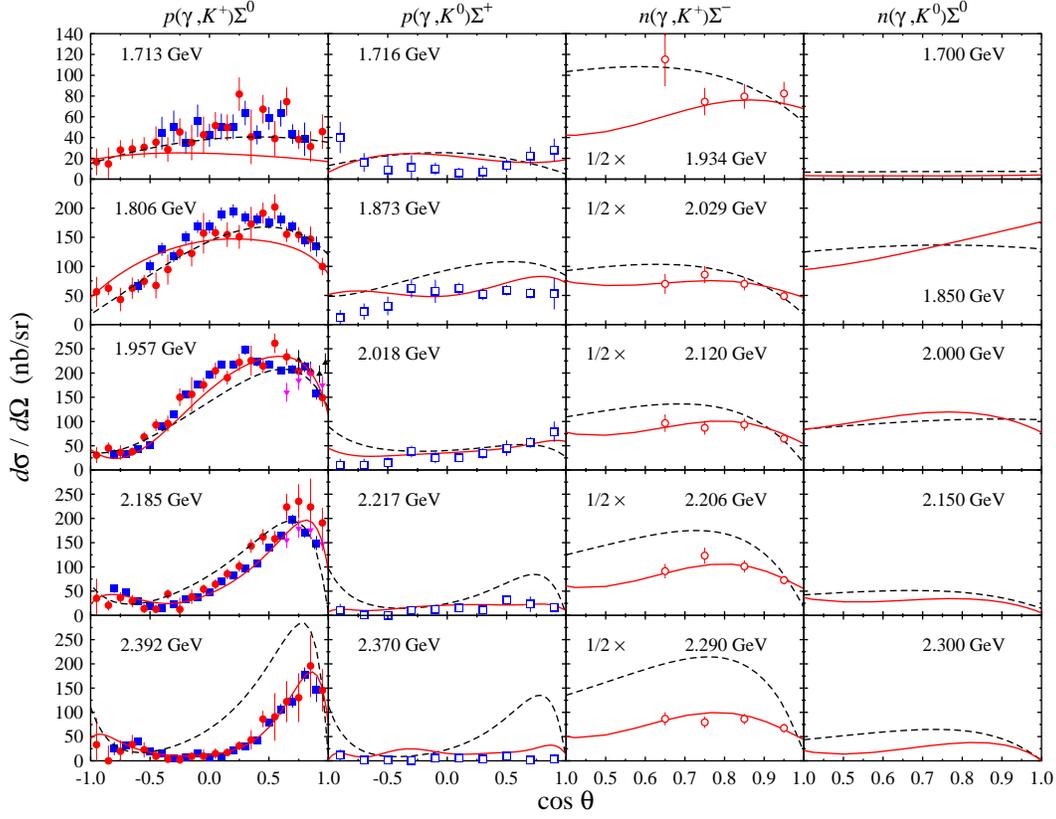}
  \caption{Comparison between the calculated differential cross sections
    from Kaon-Maid (dashed curves) and the present work (solid curves)
  with selected experimental data for all $K\Sigma$ isospin channels with selected
  energy bins. 
  Experimental data for the $K^+\Sigma^0$ channel are from 
  SAPHIR (solid circles)~\protect\cite{Glander:2003jw},
  CLAS (solid squares)~\protect\cite{Bradford:2005pt}, and LEPS (solid triangles)
  \protect\cite{Sumihama:2005er,Kohri:2006yx}. For the $K^0\Sigma^+$ channel
  experimental data are taken from SAPHIR (open squares)~\protect\cite{Lawall:2005np},
  whereas for the  $K^+\Sigma^-$ channel experimental data are from
  LEPS (open circles)~\protect\cite{Kohri:2006yx}. Note that for the
  $K^+\Sigma^-$ channel data and curves have been rescaled by a factor
  of 1/2 in order to fit on the scale.
\label{fig:dcs_sigma}}
\end{figure}

In total we use 2816 data points in our fitting data base. 
From their types the experimental data used are dominated by the differential cross
section data followed by the hyperon recoil polarization ones. From the isospin channel 
point of view, except for the $K^0\Sigma^0$ channel, all channels have experimental data. 
Most of the data were collected for the $K^+\Sigma^0$. Data for the $K^0\Sigma^+$ channel were
measured by the SAPHIR collaboration~\cite{Lawall:2005np} and are given in terms of differential cross
section and recoil polarization. For the $K^+\Sigma^-$ channel experimental data were
extracted by the LEPS~\cite{Kohri:2006yx} collaboration and are represented by differential cross section
and photon asymmetry.

For the background amplitudes we use the similar tree-level Feynman 
diagrams as in the case of $K\Lambda$. Different from the $K\Lambda$ case, 
in the $K\Sigma$ case all resonance properties, i.e.,
the mass, width, branching ratios, as well as the proton and neutron
helicity photon couplings are constrained by using the PDG values~\cite{yao:2006}.
In the fitting process we found that the $K^0\Sigma^+$ data require
a weighting factor. This is understandable, because the number
of data for the $K^0\Sigma^+$ channel 
is substantially smaller than that for the $K^+\Sigma^0$ channel, 
and the corresponding error bars are significantly larger. 
For the fit result shown in Fig.~\ref{fig:dcs_sigma}
the $K^0\Sigma^+$ channel has been weighted by a factor of 4. 
Nevertheless, as shown in this figure, 
compared to the Kaon-Maid prediction the present calculation
yields a more satisfactorily result. Predictions for the $K^0\Sigma^0$
channel is also shown in Fig.~\ref{fig:dcs_sigma}. It is obvious
that this channel is very difficult to measure. Our calculation
predicts that the corresponding cross section is comparably small as 
 the $K^0\Sigma^+$ cross section. Details of the findings in the 
$K\Sigma$ channels will be reported elsewhere~\cite{mart_future}.

\begin{theacknowledgments}
  The author acknowledges the support from the University of
  Indonesia.
\end{theacknowledgments}


\begin{thebibliography}{99}
\bibitem{kaon-maid} T.~Mart, \emph{Phys.\ Rev.\ C} {\bf 62}, 038201 (2000); C.~Bennhold,
               H.~Haberzettl and T.~Mart, arXiv:nucl-th/9909022.
\bibitem{Mart:1999ed}
  T.~Mart and C.~Bennhold, 
  \emph{Phys.\ Rev.\ C} {\bf 61}, 012201 (1999).
\bibitem{Mart:2006dk}
  T.~Mart and A.~Sulaksono,
  \emph{Phys.\ Rev.\ C} {\bf 74}, 055203 (2006).
\bibitem{hbmf}  H. Haberzettl, C. Bennhold, T. Mart, and T. Feuster,
  Phys. Rev. C {\bf 58}, R40 (1998).
\bibitem{SP98} 
  M.~Q. Tran {\it et al.}, 
  \emph{Phys. Lett. B} {\bf 445}, 20--26 (1998).
\bibitem{kaon-maid-homepage} T. Mart {\it et al.},
  http://www.kph.uni-mainz.de/MAID/kaon/kaonmaid.html.
\bibitem{Tiator:2003uu}
  L. Tiator {\it et al.}, 
  \emph{Eur.\ Phys.\ J.\ A} {\bf 19}, 55--60 (2004).
\bibitem{Glander:2003jw}
  K.~H.~Glander {\it et al.},
  \emph{Eur.\ Phys.\ J.\ A} {\bf 19}, 251--273 (2004).
\bibitem{Bradford:2005pt}
  R.~Bradford {\it et al.},
  \emph{Phys.\ Rev.\ C} {\bf 73}, 035202 (2006).
\bibitem{Sumihama:2005er}
  M.~Sumihama {\it et al.}  [LEPS Collaboration],
  \emph{Phys.\ Rev.\ C} {\bf 73}, 035214 (2006).
\bibitem{Kohri:2006yx}
  H.~Kohri {\it et al.},
  \emph{Phys.\ Rev.\ Lett.}\  {\bf 97}, 082003 (2006).
\bibitem{Bydzovsky:2006wy}
  P.~Byd\v{z}ovsk\'y and T.~Mart,
  \emph{Phys.\ Rev.\ C} {\bf 76}, 065202 (2007).
\bibitem{StatMan} E.L. Crow, F.A. Davis, M.W. Maxfield, {\it Statistics
                  Manual} (Dover Publication Inc., New York, 1960).
\bibitem{gdh} S.B. Gerasimov, Sov. J. Nucl. Phys. {\bf 2}, 430 (1966);
    S.D. Drell and A.C. Hearn, Phys. Rev. Lett. {\bf 16}, 908 (1966).
\bibitem{Mart:2008ik}
  T. Mart, arXiv:0803.0601 [nucl-th]; \emph{Few Body Syst}. {\bf 42}, 125 (2008);
  \emph{Int. J. Mod. Phys. A} {\bf 23}, 599 (2008).
\bibitem{Anisovich:2007bq}
  A.~V.~Anisovich, V.~Kleber, E.~Klempt, V.~A.~Nikonov, A.~V.~Sarantsev and U.~Thoma,
  \emph{Eur.\ Phys.\ J.\  A} {\bf 34}, 243 (2007).
\bibitem{Bradford:2006ba}
  R.~Bradford {\it et al.},
  \emph{Phys.\ Rev.\  C} {\bf 75}, 035205 (2007).
\bibitem{Mart:1995wu}
  T.~Mart, C.~Bennhold and C.~E.~Hyde-Wright,
  \emph{Phys.\ Rev.\  C} {\bf 51}, 1074 (1995).
\bibitem{Lawall:2005np}
  R.~Lawall {\it et al.},
  \emph{Eur.\ Phys.\ J.\  A} {\bf 24}, 275 (2005).
\bibitem{yao:2006} W.-M. Yao {\it et al}., \emph{J. Phys. G} {\bf 33}, 1 (2006).
\bibitem{mart_future} T. Mart, in preparation.
\end{thebibliography}
\end{document}